\documentclass[fleqn,10pt]{wlscirep}
\usepackage[utf8]{inputenc}
\usepackage{subfig}

\title{Asymmetry of short-term control of spatio-temporal gait parameters during treadmill walking}

\author[1,+]{Klaudia Kozlowska}
\author[1,+*]{Miroslaw Latka}
\author[2,+]{Bruce J. West}
\affil[1]{Wroclaw University of Science and Technology, Faculty of Fundamental Problems of Technology, Department of Biomedical Engineering, Wroclaw, 50-370, Poland}
\affil[2]{Army Research Office, Information Sciences Directorate, Research Triangle Park, 27709, USA}

\affil[*]{Miroslaw.Latka@pwr.edu.pl}

\affil[+]{each author contributed equally to this work}


\begin{abstract}

Optimization of energy cost determines average values of spatio-temporal gait parameters such as step duration, step length or step speed. However, during walking,  humans need to adapt these parameters at every step to respond to exogenous and/or endogenic perturbations. While some neurological mechanisms that trigger these responses are known, our understanding of the fundamental principles governing step-by-step adaptation remains elusive. We determined the gait parameters of 20 healthy subjects with right-foot preference during treadmill walking at speeds of 1.1, 1.4 and 1.7 m/s. We found that when the value of the gait parameter was conspicuously greater (smaller) than the mean value,  it was either followed immediately by a smaller (greater) value of the contralateral leg (interleg control),  or the deviation from the mean value decreased during the next movement of ipsilateral leg (intraleg control). The selection of step duration and the selection of step length during such transient control events were performed in unique ways. We quantified the symmetry of short-term control of gait parameters and  observed the significant dominance of the right leg in short-term control of all three parameters at higher speeds (1.4 and 1.7 m/s).

\end{abstract}
\begin{document}

\flushbottom
\maketitle

\thispagestyle
 {empty}

\section*{Introduction}

It has been known for over a century that the stride interval of human gait is remarkably stable. Small fluctuations of approximately 3\textendash{}4\%
were attributed to the complexity of the locomotor system and treated
as an uncorrelated random process \cite{Biodynamics2004}. From this
viewpoint, the discovery of long-time, fractal correlations in stride-interval
time series was unexpected \cite{Hausdorff1995,Hausdorff1996}. Those early
papers not only spurred interest in the emerging concept of fractal physiology \cite{FractalPhysiology1994}, but also shifted the focus of
quantitative gait analysis from average values of typical parameters
(e.g. stride intervals) to their temporal variability. That profound change
of perspective brought new insights into locomotor manifestations
of Huntington's and Parkinson's diseases, aging,  and the connection between gait dynamics and fall risk, see \cite{Hausdorff2007} and references therein. 

From a plethora of physiologically accessible gait patterns, humans employ only walking and running. Walking feels easiest at low speeds, and running feels easiest when moving faster. Optimization of energy cost underlies not only the choice of gait \cite{Zarrugh1974, Srinivasan2006, Srinivasan2007},  but also determines average values of gait parameters,  such as step length and duration \cite{Kito2006}. During walking,  humans need to adapt their spatio-temporal gait parameters at every step to be able to respond to exogenous (e.g. irregularities of walking surface) and/or endogenic (neuromuscular noise) perturbations \cite{Dingwell2006}. While some neurological mechanisms that trigger these responses are  known \cite{Zehr1999,Warren2001,Bent2004,Rossignol2006},  the fundamental principles governing step-by-step adaptation remain elusive \cite{Dingwell2010}.

Treadmill walking, especially at high speeds, presents challenges that can be met only through effective short-term control of spatio-temporal gait parameters. In order to stay on a treadmill, the subject's step duration and length must yield a step speed which can fluctuate over a narrow range centered on the treadmill belt's speed. The results of previous experiments with walking on a split-belt treadmill underscore the intricacies of such aggregation. In particular, spatial and temporal controls of locomotion are accessible through distinct neural circuits \cite{Reisman2005}  and neural control of intra- versus interlimb parameters (calculated using values from both legs, e.g., step length, double support) during walking is to a large extent independent \cite{Malone2010}. Herein, we investigate the dynamics of time series of gait parameters (step duration, length and speed) following a sudden, large deviation from the mean value. In particular, we test the hypothesis that whenever the value of a gait parameter is markedly greater (smaller) than the mean value, it is either immediately followed by a smaller (greater) value of the contralateral leg (interleg control), or the deviation from the mean value decreases during the next movement of ipsilateral leg (intraleg control). Said differently, during treadmill walking errors are not gradually attenuated via long-term corrections, but are corrected immediately by the same or opposite leg. Taking into account differences in the relative contribution of lower limbs to control and propulsion – the effect known as functional gait asymmetry \cite{Sadeghi1997}, we further hypothesize that in subjects with right foot preference the short-term control of gait spatio-temporal parameters is stronger for the right leg.

\section*{Results}

In Table 1 we recorded data which are crucial for testing the main hypothesis of the paper: that errors in gait parameters are  e corrected immediately by the same or opposite leg. Let us focus on the first row of this table which concerns step duration control at a treadmill speed of 1.1 m/s. There were 180 "errors" defined as abrupt changes in step duration (equations \ref{eq:Error}-\ref{eq:isolated}). For the left leg, in 42 cases (column L-L),  the deviation of step duration from the mean value did not trigger a compensating change in the step duration of the right leg. The value of the control parameter $D^{LR}$, defined by equation \ref{eq:Delta1}, greater than 1 indicates the absence of such adjustment. However, the deviation decreased during the next left step as indicated by the value of intraleg control parameter $D^{LL}$, defined by equation \ref{eq:Delta2}, smaller than 1 (the statistics of both inter- and intraleg control parameters are presented in Table \ref{tab:D-Values}). In other words, for the left leg, we observed intraleg control of step duration in 42 cases. In 51 cases (column R-L),  the change in the step duration of the left leg compensated the deviation of step duration of the previous right step (interleg control). By adding columns L-L and R-L,  we obtain 93 control events performed by the left leg. This number expressed as the percentage of all 180 control events is given in the  column of Table 1 labeled as Left. Please note that only in 4 (column NC) out of 180 cases (2.2\%),  the appearance of a step duration error did not evoke either of the control mechanisms.

\begin{table*}
\caption{The statistics of the occurrence of intraleg (columns L-L and R-R) and interleg (columns R-L and L-R) control mechanisms that are evoked in response to errors in step duration, length and velocity. Statistics are presented for three values of treadmill speed $v$. The number of errors, defined as abrupt deviations from the moving average value, is presented in the Total column. The column labeled NC gives the number of errors that persisted for more than two successive steps. The Left and Right columns show the number of control events performed by each leg. Probabilities of occurrence of intra- and interleg control mechanisms are listed in the last four columns.
\label{tab:StatisticsLR}}

\centering
\begin{tabular}{ccccccccccccc}
\hline 
$v$ {[}m/s{]} & Total & NC & L-L & R-L & R-R &  L-R  & Left {[}\%{]} & Right {[}\%{]} & $p_{LL}$ & $p_{RL}$ & $p_{RR}$ & $p_{LR}$\tabularnewline
\hline 
 &  &  &  &  &  & \multicolumn{2}{c}{} &  &  &  &  & \tabularnewline
 &  &  &  &  &  & \multicolumn{2}{c}{\textbf{step duration}} &  &  &  &  & \tabularnewline

1.1 & 180 & 4& 42 & 51 & 32 & 51  & 53  & 47 & 0.23 & 0.28 & 0.18 & 0.28 \tabularnewline
1.4 & 121 & 11 & 29 & 21 & 18 & 42 & 45 & 55 & 0.24 & 0.17 & 0.15 & 0.35 \tabularnewline
1.7 & 103 & 5 & 22 & 18 & 26 & 32 &  41 & 59 & 0.21 & 0.17 & 0.25 & 0.31 \tabularnewline
 &  &  &  &  &  & \multicolumn{2}{c}{} &  &  &  &  & \tabularnewline
 &  &  &  & \multicolumn{2}{c}{} & \multicolumn{2}{c}{\textbf{step length}} &  &  &  &  & \tabularnewline
1.1 &  380 & 14 & 125 & 71 & 122 & 48  & 54  & 46 & 0.33 & 0.19 & 0.32 & 0.13\tabularnewline
1.4 & 257 & 12 & 73 & 30 & 93 & 49  & 42 & 58 & 0.28 & 0.12 & 0.36 & 0.19\tabularnewline
1.7 & 186 & 18 & 42 & 17 & 87 & 22  &  35 & 65 & 0.23 & 0.09 & 0.47 & 0.12\tabularnewline

 &  &  &  &  &  & \multicolumn{2}{c}{} &  &  &  &  & \tabularnewline
 &  &  &  &  &  & \multicolumn{2}{c}{\textbf{step speed}} &  &  &  &  & \tabularnewline

1.1 & 393 & 5 & 83 & 115 & 84 & 106 & 51  & 49 & 0.21 & 0.29 & 0.21 & 0.27\tabularnewline
1.4 & 275 & 14 & 63 & 61 & 61 & 76 & 48 & 52 & 0.23 & 0.22 & 0.22 & 0.28\tabularnewline
1.7 & 189 & 8 & 34 & 45 & 66 & 36 &  44 & 56 & 0.18 & 0.24 & 0.35 & 0.19\tabularnewline
 &  &  &  &  &  & \multicolumn{2}{c}{} &  &  &  &  & \tabularnewline
\hline 
\end{tabular}
\end{table*}

For all three gait parameters, the number of errors decreases with speed. 
For example, there were 180 errors in step duration at $v=1.1$ $m/s$ but only 103 at $v=1.7$  $m/s$ (a 43\% reduction). A comparable drop in the number of errors was observed for step length (51\%) and velocity (52\%). Less than half of these changes can be explained by the 22\% reduction of the number of steps taken by all the subjects at $v=1.7$ $m/s$ in comparison with $v=1.1$ $m/s$. Please note that the number of steps decreases with treadmill speed since at each speed,  the subjects were asked to cover the same distance of 400 m. It is worth emphasizing that there were approximately twice as many errors in step length and speed than in step duration. For all three parameters: step duration, length, and speed, in at least 90\% of cases,  the  deviations from the mean value decreased during the subsequent two steps via either intra- or interleg control.

Regardless of treadmill belt speed,  the control of step length is predominately intraleg (Table \ref{tab:StatisticsLR}).  For example, for the left leg at the lowest speed,  the probability of evoking intraleg control $p_{LL}=0.33$ is 42\% greater than that of interleg control $p_{RL}=0.19$). In the same condition, for the right leg  such difference is equal to 59\% ( $p_{RR}=0.32$ vs $p_{LR}=0.13$).
There is no such pattern for the other two gait parameters.

For step duration, length, and speed, the control parameter $D$ was independent of speed (Table \ref{tab:D-Values}). For all three gait parameters, both for the right and left leg, the mean value of $D$ for interleg control was greater than that of intraleg control. For example, for step duration at $v=1.1 m/s$  $D_{RL}=0.55$ and  $D_{LL}=0.34$.

The difference between the values of intra- and interleg control parameters for a given leg was statistically significant for all three treadmill speeds 
for step length:
\begin{itemize}
\item at 1.1 m/s:  $p_{left} < 1 \times 10^{-4}$, $p_{right} = 6 \times 10^{-3}$;
\item at 1.4 m/s:  $p_{left} < 1 \times 10^{-4}$, $p_{right} = 3 \times 10^{-2}$;
\item at 1.7 m/s:  $p_{left} = 1 \times 10^{-4}$, $p_{right} = 2 \times 10^{-3}$;
\end{itemize}
as well as the step speed:
\begin{itemize}
\item at 1.1 m/s:  $p_{left} < 1 \times 10^{-4}$, $p_{right} = 6 \times 10^{-4}$;
\item at 1.4 m/s:  $p_{left} < 1 \times 10^{-4}$, $p_{right} < 1 \times 10^{-4}$;
\item at 1.7 m/s:  $p_{left} = 2 \times 10^{-3}$, $p_{right}  = 9 \times 10^{-3}$.
\end{itemize}
For step duration such differences were not so strongly pronounced:
\begin{itemize}
\item at 1.1 m/s:  $p_{left} = 5 \times 10^{-4}$, $p_{right} = 5 \times 10^{-2}$;
\item at 1.7 m/s:  $p_{left} = 2 \times 10^{-3}$.
\end{itemize}

With the exception of step duration and step length at the lowest speed ($v=1.1$ $m/s$), the asymmetry parameter $\Delta DBE$ was smaller than zero indicating a dominant role of the right leg in short-term control of gait parameters during treadmill walking.

\begin{table*}
\caption{The values of intra- and interleg control parameters for treadmill walking. Data are presented as mean  (standard deviation).\label{tab:D-Values}}

\centering
\begin{tabular}{ccccccccc}
\hline 
&  &  &  &   \multicolumn{2}{c}{} &  &   \tabularnewline
$v$ {[}m/s{]} & $D^{LL}$ & $D^{RL}$ & $D^{RR}$ & $D^{LR}$ & $DBE_{L}$ & $DBE_{R}$ & $\Delta DBE$ {[}\%{]} & \tabularnewline
\hline 
&  &  &  &   \multicolumn{2}{c}{} &  &   \tabularnewline
 &  & \multicolumn{4}{c}{\textbf{step duration}} &  &  & \tabularnewline

1.1 & 0.34 (0.22) & 0.55 (0.26) & 0.43 (0.25) & 0.58 (0.21) & 1.20 & 0.91 & 28 & \tabularnewline
1.4 & 0.45 (0.26) & 0.55 (0.25) & 0.31 (0.22) & 0.54 (0.24) & 0.85 & 1.12 & -28 & \tabularnewline
1.7 & 0.31 (0.25) & 0.63 (0.20) & 0.36 (0.28) & 0.49 (0.26) & 0.97 & 1.34 & -32 & \tabularnewline
&  &  &  &   \multicolumn{2}{c}{} &  &   \tabularnewline
 &  & \multicolumn{4}{c}{\textbf{step length}} &  &  & \tabularnewline
 1.1 & 0.46 (0.26) & 0.72 (0.23) & 0.48 (0.25) & 0.62 (0.30) & 0.98 & 0.88 & 11 & \tabularnewline
1.4 & 0.46 (0.27) & 0.79 (0.16) & 0.45 (0.26) & 0.61 (0.30) & 0.76 & 1.12 & -38 & \tabularnewline
1.7 & 0.44 (0.26) & 0.79 (0.15) & 0.44 (0.27) & 0.68 (0.23) & 0.63 & 1.24 & -65 & \tabularnewline
&  &  &  &   \multicolumn{2}{c}{} &  &   \tabularnewline
 &  & \multicolumn{4}{c}{\textbf{step speed}} &  &  & \tabularnewline
1.1 & 0.44 (0.26) & 0.65 (0.26) & 0.40 (0.27) & 0.56 (0.28) & 0.92 & 1.01 & -9 & \tabularnewline
1.4 & 0.39 (0.25) & 0.68 (0.24) & 0.33 (0.23) & 0.56 (0.29) & 0.91 & 1.15 & -23 & \tabularnewline
1.7 & 0.51 (0.26) & 0.73 (0.20) & 0.41 (0.24) & 0.58 (0.27) & 0.68 & 1.19 & -55 & \tabularnewline
&  &  &  &   \multicolumn{2}{c}{} &  &   \tabularnewline

\hline 
\end{tabular}
\end{table*}

Table \ref{tab:Compensation} shows the probability of compensatory response to errors in gait spatio-temporal parameters for intra-  (L-L, R-R)  and interleg (L-R, R-L) control. Such response corresponds to negative values of variables $S^{inter}$ (equation \ref{eqn:SignInter}) and $S^{intra}$ (equation \ref{eqn:SignIntra}). For all speeds and parameters, the probability of interleg compensation is close to 1, roughly two times higher than that of intraleg response.

\begin{table}
\caption{Probability of compensation of errors in gait spatio-temporal parameters for intra-  (L-L, R-R)  and interleg (L-R, R-L) control. Statistics are presented for three values of treadmill speed $v$. 
\label{tab:Compensation}}

\centering
\begin{tabular}{ccccccccccccc}

\hline 
$v$ {[}m/s{]}  & L-L & R-R & L-R &  R-L    & L-L & R-R & L-R &  R-L
 & L-L & R-R & L-R &  R-L
\tabularnewline
\hline 

\tabularnewline
   
  &
  \multicolumn{4}{c}{\bf{step duration}} &
  
  \multicolumn{4}{c}{\bf{step length}} &
  
  \multicolumn{4}{c}{\bf{step speed}}

  \tabularnewline

1.1 & 0.50& 0.42 & 0.98 & 0.94  & 0.40& 0.38 & 0.83 & 0.83 & 0.52& 0.55 & 0.92 & 0.92  \tabularnewline

1.4 & 0.28& 0.39 & 0.95 & 0.95  & 0.42& 0.34 & 0.82 & 0.80 & 0.52& 0.46 & 0.93 & 0.92  \tabularnewline

1.7 & 0.50& 0.54 & 0.97 & 0.94  & 0.40& 0.43 & 0.95& 0.76 & 0.32& 0.44 & 0.95 & 0.89  \tabularnewline

 \\
\hline 
\end{tabular}
\end{table}

\vspace*{3\baselineskip}

\section*{Discussion}

In overground walking with self-selected speed, fluctuations of stride interval, length, and speed exhibit persistent fractal scaling characterized by a Hurst exponent $\alpha > 0.5$  \cite{Hausdorff1995,Hausdorff1996,Terrier2005}. Auditory metronomic cueing changes fractal statistics of stride intervals from persistent
to antipersistent ($\alpha < 0.5$) \cite{Delignieres2009}. The super central pattern generator model, introduced  by West and Scafeta \cite{PhysRevE.67.051917}, elucidates the dynamic origin of such transitions. In particular, the transitions result from the driving of a fractal clock,  which retains its properties under perturbation. In treadmill walking, fluctuations of interstride interval and stride length are also persistent. However, the time series of stride speed is antipersistent,  which is a manifestion of increased central control of this gait parameter \cite{Dingwell2010,Terrier2012}. Terrier has recently demonstrated that visual cueing (alignment of step lengths with marks on the floor) also induced anti-correlated pattern  in gait parameters \cite{Terrier2016}.

To a large extent, fluctuations of spatio-temporal gait parameters result from the intrinsic fractal properties of pattern generators. Hidden in these fluctuations are sporadic control events, triggered to accomplish a locomotor task such as remaining on a moving treadmill belt. This is why we study the dynamics of time series of gait parameters that follow a sudden large deviation from a mean value. For lack of a better word, we dubbed such events errors, but emphasize that they may originate either from the failure of the motor control system, or from the necessary adjustment of the subject's position on a treadmill.  While the definition of such events is arbitrary (equations \ref{eq:Error}-\ref{eq:isolated}),  it satisfies the research objective.

We found that when the value of the gait parameter (step duration, length or speed) was conspicuously greater (smaller) than the mean value, it was either followed immediately by a smaller (greater) value of the contralateral leg (interleg control), or the deviation from the mean value decreased during the next movement of ipsilateral leg (intraleg control). The existence of distinct short-term control of step frequency (the inverse of step duration) was demonstrated by Snaterse et al. \cite{Snaterse2011}. The time evolution of step frequency triggered by sudden stepwise increments in treadmill speed was modeled by the sum of two exponentially decaying terms. The time constant of the first term was $1.44 \pm 1.14$ s and its amplitude was two times larger than that of the second term, whose time constant was $27.56 \pm 16.18 s$. For those values of time constants,  step frequency adjustments were two-thirds complete in less than two seconds. Snaterse et al. argued that the first term represents a rapid pre-programmed response, while the slower one models fine-tuning of step frequency driven by energy expenditure optimization. Herein we extended this line of reasoning by demonstrating that short-term control of gait parameters may be realized using intra- and interleg adjustments. The better understanding of short-term control mechanisms does not  bring us any closer to understanding how, during treadmill walking, persistent stochastic variables: step duration and step length are combined to yield antipersistent step speed. We believe that a different mechanism  operating at a longer time scale underlies this effect. 

There are fundamental differences between the control of step duration and step length. The probability of evoking intraleg control of step length at the highest treadmill speed ( $v=1.7 m/s$) is approximately three times greater than that of evoking interleg control. There is no such distinct pattern for step duration. Moreover, the number of errors in step duration is half that of step length, regardless of treadmill belt speed. This is a strong indication that spatial and temporal controls of locomotion are accessible through distinct neural circuits. This interpretation is corroborated by the earlier study of Malone and Bastian, who investigated adaptation of spatial and temporal aspects of walking to a sustained perturbation, generated by a split-belt treadmill \cite{Malone2010}. They demonstrated that conscious correction facilitates adaptation, whereas distraction slows it. The unexpected finding of their study was that those manipulations affected the adaptation rate of the spatial elements of walking, but not of the temporal ones. In the follow-up study Malone et al \cite{Malone2012} demonstrated that temporal and spatial controls of symmetric gait can be adapted independently. Please note that continuous, conscious assessment of distance to surrounding objects lies at the heart of the control problem of remaining stationary on a moving treadmill belt. Thus, the large number of errors in step length as compared to step duration may reflect both the dominant role of spatial control and its susceptibility to distraction. It is worth mentioning that in casual walking, the coefficient of variation of stride time is much smaller than that of stride length and of walking speed \cite{Kito2006}.

Step speed may be interpreted as the output of the intricate neuromuscular control system, which integrates different sensory-motor processes. The ratio of average values of step length and frequency, or walk ratio, is constant over a broad range of walking speeds. In other words, there is a linear relation between these gait parameters (the stride length -- cadence relationship), a pre-programmed pattern which presumably simplifies gait control in steady state walking \cite{Egerton2011}. Let us analyze the interplay of step duration and step length during transient changes following the occurrence of errors. We previously pointed out that these two parameters are controlled in distinct ways. In particular, the probability of evoking the interleg control of step length is at least two times smaller than that of evoking the intraleg control (Table \ref{tab:StatisticsLR}). In sharp contrast, the probability of either  inter- or  intraleg control of step speeds is comparable. Thus, we may hypothesize that negative-feedback adjustment of step duration of the contralateral leg underlies the interleg control of step speed. It is worth emphasizing that the intraleg control of step speed is stronger than the interleg control. 

The recent work of Dingwell et al. \cite{Dingwell2010} provides insight into the maintenance of speed during treadmill walking. A subject can in principle choose any combination of stride length and time that yields step speed equal to that of a treadmill belt. These pairs of values form in phase-space a diagonal line called a goal equivalent manifold (GEM) \cite{Cusumano2006}. Dingwell et al. decomposed deviation from this manifold into tangent and transverse components. Only the latter component was tightly controlled. Moreover, the time series of transverse deviations exhibited statistical antipersistence characteristic of stride speed. This study underscores the significance of interleg control of gait parameters. We believe that the GEM decomposition should be applied to time series of step velocities to quantify the interleg control in a more sophisticated way.

In able-bodied gait, asymmetry in spatio-temporal and kinematic parameters (such as speed profiles, step and stride length, foot placement angle, maximum knee flexion) for the left and right leg has been frequently reported \cite{Sadeghi2000}. To the best of our knowledge, the present study is the first observation of asymmetry in dynamics of human gait parameters. With the exception of step duration control at the lowest speed, for all three gait parameters $\Delta DBE < 0$, indicating dominance of the right leg in short-term control. The origin of this asymmetry can be traced back to differences in the relative contribution of lower limbs to control and propulsion – the effect known as functional gait asymmetry \cite{Sadeghi1997}. More specifically, the leg with greater muscle power generation dominates propulsion, while the support and control functions are more conspicuous for the leg with greater power absorption.  Humans are typically right-footed for mobilization and left-footed for postural stabilization.

Special consideration should be given to step duration and step length control at the lowest speed $v=1.1 m/s$. Only in this case, the asymmetry parameter $\Delta DBE$ was greater than zero, indicating the dominance  of left lower limb. Note that the lowest asymmetry, $|\Delta DBE|$, was observed for all three parameters at $v=1.1 m/s$. Differences in low-speed gait have been reported before. Terrier and Schutz \cite{Terrier2003} demonstrated that during overground walking, at low speeds the majority of subjects adopted a higher walk ratio and had a higher variability of stride time. However, in this study the lowest treadmill speed coincides with the preferred walking speed (PWS) of young subjects \cite{Terrier2012}. There are two possible explanations for the positive value of $\Delta DBE$. It is likely that in the vicinity of PWS priority is given to balance maintenance and consequently stride duration control is shifted to the left leg, which is used for postural stabilization. Please note that our cohort included only subjects with clearly pronounced right foot preference. Alternatively, reversed asymmetry for step duration and low values of $|\Delta DBE|$ for step length and step speed  may indicate that there exists a different strategy for control of gait parameters in overground walking (treadmill walking at $v=1.1 m/s$ may not be challenging for young subjects and may resemble unconstrained overground walking). This argument is plausible because in motor coordination tasks, humans correct only those deviations that interfere with task goals and allow variability in redundant (task-irrelevant) dimensions \cite{Todorov2002}. Following the logic of this minimum intervention principle, in treadmill walking, step speed must be tightly regulated. However, in overground walking, higher priority may be given, for example, to balance control, which would affect the value of the asymmetry parameter $\Delta DBE$. These two qualitatively different strategies may also reflect other fundamental differences between overground and treadmill walking. The rate at which the environment flows past the eyes seems to be an important mechanism for regulating walking speed \cite{Mohler2007,OConnor2012}. More specifically, vision is used correctively to maintain walking speed at a value that is perceived to be optimal. For treadmill walking, a discrepancy between observed and expected visual flow leads to a significant reduction (about 20\%) of PWS \cite{Dal2010}, as well as the speeds of walk-run and run-walk transitions \cite{Mohler2007}. It is worth pointing out that as far as kinetic and kinematic parameters are concerned, treadmill and outdoor gaits are similar \cite{Riley2008}.  

The discovery of dependence  of functional asymmetry in short-term control of gait spatio-temporal parameters on treadmill speed was an unexpected outcome of this research. The elucidation of the transition from left-leg to right-leg dominance in short-term control entails determination of the PWS for each subject. Further research is also needed to understand why the probability of compensatory response for interleg control is close to 1 and is almost two times greater than that of intraleg control (Table \ref{tab:Compensation}). Undoubtedly, such a strong difference indicates different roles  these two mechanism play in control of gait during treadmill walking.  One may hypothesize that the primary goal of interleg control is maintenance of balance via negative feedback from either leg while achieving specific goals such as matching the speed of the treadmill belt requires intraleg adjustments.

During human locomotion, the legs act as two coupled oscillators \cite{Beek1995}. However, most studies disregard bilateral coordination and synchronization dynamics \cite{Bartsch2007,Krasovsky2012,Wuehr2014} and focus on single-leg variability (stride time, length, speed). Herein we demonstrated asymmetric short-term intra- and interleg control of spatio-temporal gait  parameters. We believe that a better understanding of these effects will not only pave the way for more realistic models of gait variability and control, but also help to refine procedures used in rehabilitation of gait impairments.

\section*{Methods}

We recruited 20 healthy students (10M/10F, mean(SD): age 22 yr (2), height 1.73 m (0.1), weight 71 (15) kg, BMI 23 (4)) of the Wroclaw University of Science and Technology, who all signed an informed consent. The study was performed according to the Declaration of Helsinki and the protocol was approved by the Ethics Committee of Wroclaw Medical University. The subjects were screened to exclude those with a history of orthopedic problems, recent lower extremity injuries, any visible gait anomalies, or who were taking medications that might have influenced their gait. We only enrolled subjects who used the right leg to: kick a tennis ball, manipulate a tennis ball around a circle, make a first step, make a step after being pushed from behind. These purely bilateral tasks are frequently incorporated into foot-preference inventories \cite{Chapman1987,Gabbard1996}. The  protocol began with a 5 min familiarization period of walking on a level motor-driven treadmill. Then each subject was asked to walk 400 m three times  at 1.1 m/s, 1.4 m/s i 1.7 m/s (4 km/h, 5 km/h and 6 km/h). The objective  was to investigate control of gait parameters at treadmill speeds equal to  or greater than the PWS of young subjects. Therefore, the lowest speed was equal to the preferred walking speed reported by Terrier and Deriaz \cite{Terrier2012} and slightly smaller than the  values determined by Dal et al. \cite{Dal2010} (1.19 m/s) and Dingwell \cite{Dingwell2006} (1.22 m/s).

The gait parameters were extracted from the trajectories of the 30 mm optical markers attached to both shoes below the ankle. The movements of those markers were recorded using an in-house motion capture system with a frame rate of 240 Hz and 720p resolution. The optical tracking was implemented in C++ (Visual Studio 2013) using OpenCV library. A heel strike was defined as the point where the marker of the forward foot was at its most forward point during each gait cycle. A step length was the distance between the ipsilateral and contralateral heel strikes. A step duration was equal to the elapsed time between the ipsilateral and contralateral heel strikes. 
A step speed was calculated as the quotient of step length and step duration. The group averaged number of steps taken per trial was equal to 593 (23) at 1.1 m/s, 508 (59) at 1.4m/s, and 456 (56) at 1.7 m/s.

\begin{figure}[h!t]
\centering

\includegraphics[scale=1.00]{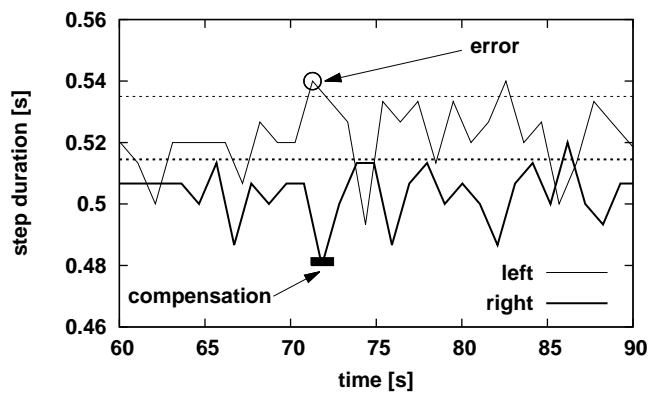}
\caption{\label{fig:TimeSeries} Time series of step durations for treadmill walking at 4 km/h. The circle indicates the step duration of left leg which was longer than the mean value (represented in this figure by horizontal, thick, dotted line) by more than 3 ⁄ 2 of standard deviation (the upper, horizontal, thin dotted gridline represents this threshold). It is apparent that the duration of the step which immediately follows the “error” suddenly decreases.}

\end{figure}

In Fig. \ref{fig:TimeSeries} we present a time series of step duration
for treadmill walking at 1.1 m/s. The circle in this figure indicates
step duration that was longer than the mean value (represented in
this figure by horizontal, thick, dotted line) by more than $3/2$
of standard deviation (the upper, horizontal, thin dotted gridline
represents this threshold). It is apparent that the duration of the
step, which immediately follows the "error", suddenly decreases
(this shorter interval is marked by the filled rectangle). This example
hints at the existence of an interleg control mechanism that stabilizes the stride
interval.

Let $N$ be the number of steps taken by each leg. Let us introduce a notation that facilitates the analysis of interleg control. We write the time series of length $2N$ of one of the gait parameters (step duration, length or speed) $\left\{ S_{n}\right\} _{n=1}^{2N}$
in the following form:
\begin{equation}
\left\{ S_{n}\right\} =\left\{ I_{1},C_{1},I_{2},C_{2}...I_{N},C_{N}\right\} ,\label{eq:Series}
\end{equation}
where subseries$\left\{ I_{j}\right\} _{j=1}^{N}$ and $\left\{ C_{j}\right\} _{j=1}^{N}$  correspond
to the ipsilateral and the contralateral leg, respectively. $\sigma_{I}$ and $\sigma_{C}$ are standard deviations of these series. The simple moving averages (the unweighted mean of the previous $m$ data) of  $\left\{ I_{j}\right\}$ and $\left\{ C_{j}\right\}$ are denoted  by $\bar{I}(m)$ and $\bar{C}(m)$, respectively.

 We define as errors
these values  $I_{i}$ which satisfy all of the following criteria:
\begin{eqnarray}
\left|I_{i}-\bar{I}(m)\right| & > & 1.5\sigma_{I}\label{eq:Error}\\
100\%\left|\frac{I_{i}-I_{i-1}}{I_{i-1}}\right| & > & 3\%\label{eq:Abrupt}\\
\left|C_{i-1}-\bar{C}(m)\right| & < & 0.5\sigma_{C}\label{eq:isolated}
\end{eqnarray}
These undoubtedly heuristic criteria are used to detect abrupt changes (equation (\ref{eq:Abrupt})) which lead to conspicuous deviations from the moving average value (equation (\ref{eq:Error}))
and which are not brought about by a deviation in the preceding step
of the contralateral leg (equation (\ref{eq:isolated})). As previously mentioned, we dub such events errors, but bear in mind that they may originate either from the motor control system failure, or from the necessary adjustment of the subject's position on a treadmill. The rationale for using the moving average in the above definition of an error stems from non-stationarity of gait time series. This modification ensures that during transient linear trends the large deviation from the global mean value does not invoke the detection algorithm. Please note that equation (\ref{eq:Abrupt}) by itself is another safeguard for false error detection caused by the transient drift of local mean value. Herein, we report the values for $m=10$.

\begin{figure}[h!t]
\includegraphics[scale=0.85]{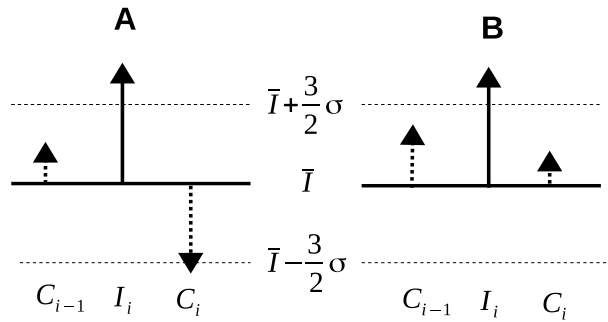}
\centering

\caption{\label{fig:DInter}
In interleg control the gait parameter of contralateral leg $C_{i}$ changes in such a way as to decrease  deviation of $I_{i}+C_{i}$ from the mean value. a) 
The error of ipsilateral leg $I_{i}$ is immediately compensated for by the contralateral leg. b) In the alternative scenario, only the magnitude of the deviation of the contralateral leg from the mean value decreases.
}
\end{figure}

Let us use $\Delta$ to denote a deviation of a given gait parameter from its moving average value, e.g. $\Delta I_{i}=I_{i}-\bar{I}(m)$. In\emph{ interleg} \emph{control} the gait parameter of contralateral leg $C_{i}$ changes in such a way as to decrease  deviation of $I_{i}+C_{i}$. To quantify such stabilization, we introduce the following metric:
\begin{equation}
D_{i}^{inter}=\frac{\left|\Delta I_{i}+\Delta C_{i}\right|}{\left|\Delta I_{i}+\Delta C_{i-1}\right|}.\label{eq:Delta1}
\end{equation}
The stabilization occurs when $D_{i}^{inter}<1$. The numerator in the above equation may become smaller than the denominator in two cases. In the first case the $\Delta C_{i}$ has the opposite sign to $\Delta I_{i}$:
\begin{equation}
S_{i}^{inter}= \Delta I_{i} \Delta C_{i} < 0,
\label{eqn:SignInter}
\end{equation} in other words, the contralateral leg \textit{compensates} for errors, as shown in Fig. \ref{fig:DInter}a. The perfect compensation corresponds to $D_{i}^{inter}=0$. In the alternative scenario, only the magnitude of the deviation of the contralateral leg from the mean value decreases ( $S_{i}^{inter}>0$) as illustrated by Fig. \ref{fig:DInter}b.

\begin{figure}[h!t]
\includegraphics[scale=0.85]{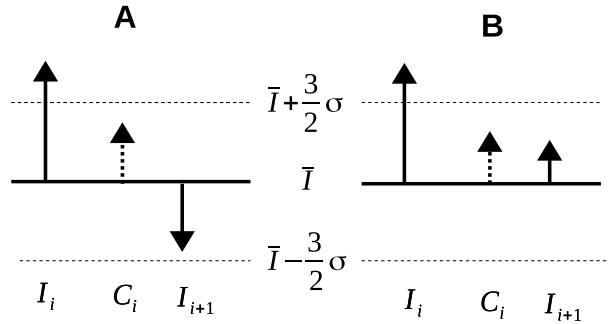}
\centering
\caption{\label{fig:DIntra} The error in a gait spatio-temporal parameter may  not evoke a sudden change of the parameter of contralateral leg. However, the deviation of the  next ipsilateral gait parameter $I_{i+1}$ from the mean value $\bar{I}$ may decrease as a result of: a) compensation or b) the reduction of the magnitude of the displacement of parameter of the ispislateral leg from the mean value.
}
\end{figure}

It is possible that an error does not bring about a sudden change of gait parameter of contralateral leg. In this case,  $D_{i}^{inter}>=1$. However, stabilization may occur during the next step of ipsilateral leg. The change of $I_{i+1}$ may reduce the deviation $\Delta I_{i+1} + \Delta C_{i}$. We refer to such a scenario as an \emph{intraleg control} and define a corresponding metric:
\begin{equation}
D_{i}^{intra}=\frac{\left|\Delta I_{i+1}+\Delta C_{i}\right|}{\left|\Delta I_{i}+\Delta C_{i}\right|}.\label{eq:Delta2}
\end{equation}
To be able to directly compare the properties of both types of control (inter- and intraleg) we distinguish whether the intraleg control was achieved via compensation:
\begin{equation}
S_{i}^{intra}= \Delta I_{i} \Delta I_{i+1} < 0,
\label{eqn:SignIntra}
\end{equation}
as shown in Fig. \ref{fig:DIntra}a, or by the reduction of the magnitude of the displacement of gait parameter of the ispislateral leg from the moving average value (Fig. \ref{fig:DIntra}b).

\begin{figure}[h!t]
\includegraphics[scale=0.85]{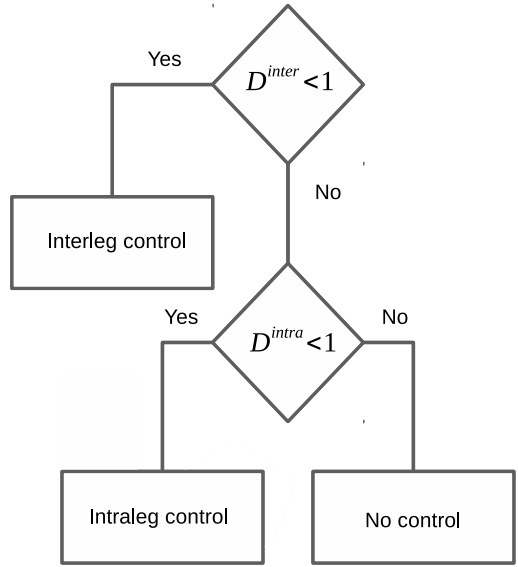}
\centering
\caption{\label{fig:Flowchart-of-analysis}Flowchart of analysis of dynamics
of gait parameter which follows an error -- gait-parameter's value which satisfy
the criteria described by equations (\ref{eq:Error}-\ref{eq:isolated}).}
\end{figure}

The flowchart in Fig. \ref{fig:Flowchart-of-analysis} elucidates the
analysis of the dynamics of gait parameter time series which follows
the occurrence of errors. Using $D^{inter}$ and $D^{intra}$, we detect
the activation of inter- and intraleg control mechanism, respectively.

In our analysis of the experimental data, we use a more specific notation for the interleg parameter $D^{inter}$. For example, to indicate that
an error in a given gait parameter of the left leg was followed by an adjustment of this parameter by the right leg,  we write $D^{LR}$. In the same vein, we use $D^{LL}$ , $D^{RR}$ to denote intraleg leg control parameter for the left and right leg, respectively. 

In most cases $D$ values, for a given gait parameter, speed, and control type, were not normally distributed (the Shapiro-Wilk test). For a given speed and gait parameter, the Levene's test showed equality of variances among the control types (with the exception of step duration at 1.1 m/s and step length at all speeds). For a given  gait parameter and control type (L-L, R-R, L-R, R-L), we  investigated the dependence of $D$ on treadmill speed. In this case, the Levene's test showed homogeneity of variance. Consequently, the Kruskal-Wallis test with Tukey’s post hoc comparisons was used to detect differences across speed and control type.  The significance threshold was set to 0.05.

To quantify functional asymmetry in control of gait spatio-temporal parameters  we need to take into account the stochastic aspect of motor control system.  Let us employ an analogy of detailed balance equation of statistical physics \cite{Kampen2007} 
and call it gait detailed balance equation (DBE). In its original formulation, detailed balancing relates the relative population of two states by the probability of a transition between them. The principle applies equally well to physical systems, mathematical probability densities, or statistical processes in a variety of forms.

The smaller $D$ the better stabilization of stride gait parameters.
Consequently,  the influence of a control mechanism (inter or intra) on gait parameters is proportional
to its probability of occurrence and the inverse of the corresponding
mean value of control parameter $\bar{D}$. For example, for the right lower limb, we may write:
\begin{equation}
DBE_{R}=\frac{p_{RR}}{\bar{D}^{RR}}+\frac{p_{LR}}{\bar{D}^{LR}}, \label{eq:DBER}
\end{equation}
in the same vein, for the left lower limb:
\begin{equation}
DBE_{L}=\frac{p_{LL}}{\bar{D}^{LL}}+\frac{p_{RL}}{\bar{D}^{RL}}.\label{eq:DBEL}
\end{equation}
The perfect symmetry corresponds to the following equality:
\begin{equation}
DBE_{L}=DBE_{R}. \label{eq:DBE}
\end{equation}
We quantify the asymmetry in control of gait spatio-temporal parameters with the relative difference expression:
\begin{equation}
\Delta DBE=2\frac{DBE_{L}-DBE_{R}}{DBE_{L}+DBE_{R}}.\label{eq:DeltaDBE}
\end{equation}

\section*{Author contributions statement}

M.L and B.J.W conceived the experiment, K.K. conducted the experiment, all authors analyzed the results and contributed to the manuscript. 

\section*{Additional information}

\textbf{Competing financial interests} There were no conflicts of interest.

\newpage

\end{document}